\documentclass[aps,showpacs,floats]{revtex4}
\usepackage{amssymb}
\usepackage{graphicx}
\usepackage{bm}

\begin{document}

\def\v#1{{\bf #1}}
\newcommand{\la}{\langle}
\newcommand{\ra}{\rangle}
\newcommand{\lc}{\lowercase}

\title{Metal-insulator transition in the quarter- filled frustrated checkerboard lattice}
\author{Y. Z. Zhang$^1$, Tran Minh-Tien$^{1,3}$, V. Yushankhai$^{1,4}$, and P. Thalmeier$^2$}
\affiliation{$^1$Max-Planck-Institut f\"{u}r Physik komplexer Systeme, N\"{o}thnitzer
Stra\ss e 38 01187 Dresden, Germany\\
$^2$Max-Planck-Institut f\"{u}r Chemische Physik fester Stoffe,
N\"{o}thnitzer Stra\ss e 40 01187 Dresden, Germany\\
$^3$Institute of Physics, P. O. Box 429, Bobo, 10000 Hanoi, Vietnam\\
$^4$Joint Institute for Nuclear Research, Dubna, Russia}

\bibliographystyle{apsrev}

\begin{abstract}
We study the electronic structure and correlations in the geometrically
frustrated two dimensional checkerboard lattice. In the large U limit
considered here we start from an extended Hubbard model of spinless fermions at
half-filling. We investigate the model within two distinct Green's function
approaches: In the first approach a single-site representation decoupling
scheme is used that includes the effect of nearest neighbor charge fluctuations.
In the second approach a cluster representation leading to a
'multiorbital' model is investigated which includes intra-cluster
correlations exactly and those between clusters on a mean field basis. It is
demonstrated that with increasing nearest-neighbor Coulomb interaction V
both approaches lead to a metal-insulator transition with an associated
'Mott-Hubbard' like gap caused by V. Within the single site approach we also
explore the possibility of charge order. Furthermore we investigate the
evolution of the quasiparticle bands as funtion of V.
\end{abstract}


\maketitle

\section{Introduction}
To treat the problem of a metal-insulator transition (MIT) driven by
electron-electron interactions in lattice models with a fractional
electron site occupancy,  both the on-site and long-range Coulomb repulsions
are equally important \cite{Imada}. A lattice with geometrical frustration 
imposes additional complications and may produce new features
of MIT. The well known  example is the Verwey transition
\cite{Verwey} in magnetite ($Fe_3O_4$) at $T_V \approx$
120K. Magnetite has a spinel 
structure and $A$ sites are occupied regularly with  $Fe^{3+}$
ions. On the $B$ sites forming a pyrochlore lattice, the $Fe$ ions
are in the mixed valent  state of  $Fe^{2+}:Fe^{3+}$=1:1. Therefore,
in the high-temperature charge disordered state, $T>T_V$, a band structure
calculation would predict a quarter-filled  conduction band, i.e., 
a metallic state with one itinerant electron per two sites. Below the Curie
temperature $T_C \approx 850K$ of magnetite the  itinerant
electrons are (almost) fully ferromagnetically  polarized. Therefore they can 
be treated as a system of interacting spinless fermions. In
terms of spinless fermions, the charge occupancy of one electron per
two sites corresponds to a half band-filling. Below
the Verwey temperature $T_V$, the charge degrees of freedom are
ordered and an insulating state occurs. As was first pointed out by Anderson
\cite{Anderson}, properties of the Verwey transition can not be
understood without taking into account the geometrical frustration
and the resulting huge degeneracy of the ground-state charge
configurations in the pyrochlore lattice. Two more recent examples of
electronic structure changes connected to charge ordering in spinels
are found in  $AlV_2O_4$ with temperature decreasing at ambient
pressure \cite{Matsuno,Zhang1} and in $LiV_2O_4$ under the external pressure
\cite{Takagi,Zhang1}. An interesting observation is that charge
ordering  in these three spinel
systems is accompanied with a lattice structural change,  i.e., the
system tries to avoid the geometrical frustration of the pyrochlore structure.
At the same time, no MIT and  charge ordering were observed in
$LiV_2O_4$ at ambient pressure down to very low temperatures, but
instead the metallic compound $LiV_2O_4$ exhibits a heavy-fermion
behavior below 30K \cite{Kondo,Urano}.


 Most of the
work until now has been devoted to an understanding of the magnetic
properties of geometrically frustrated lattices such as pyrochlore
structure, since in the presence of antiferromagnetic interaction
frustration acts against a conventional long-range order and may stabilize a
spin-liquid state \cite{Magnetism}. Charge degrees of freedom have in
contrast been studied much less \cite{Fulde}. In fact, charge ordering in
geometrically frustrated systems has been an intriguing and unsettled problems \cite{Fujimoto}. 
For example, numerical diagonalizations of a Hamiltonian for
spinless fermion with strong nearest-neighbor repulsion on a checkerboard
lattice have given evidence that at half filling (number of spinless fermion
equals one half the number of sites) the ground state is two-fold degenerate
but a liquid \cite{Runge}. This must be kept in mind when that system is
treated within different approximations leading to a charge ordered ground
state. In other words, the observed charge order in frustrated structures
could crucially depend on associated lattice deformations, i.e., the
involvement of lattice degree of freedom.

The extended Hubbard model on non-frustrated lattices has been extensively investigated 
in one dimension at quarter-\cite{Penc} or half-filling\cite{Zhang}, for two-leg
ladders at quarter-filling\cite{Vojta}, for two-dimensional square lattices
at half-filling\cite{YZhang} and in the limit of infinite dimensions at
quarter-\cite{Pietig} or half-filling\cite{Dongen}. A variety of techniques,
such as Hartree-Fock approximation, pertubation theory, dynamical mean-field
theory, as well as numerical methods, e.g., quantum Monte Carlo and
density-matrix renormalization group have been employed. However all these
investigations were based on the non-frustrated lattice.

In the present paper, we apply a Green's function approach to study the
possible phase transitions of a half-filled spinless fermion model , i.e., a
system with one electron per two sites, on the frustrated checkerboard
lattice (see the inset of Fig. 1). The lattice can be viewed as a
two-dimensional projection of a pyrochlore lattice. The spinless fermion
model arises naturally for ferromagnetic materials in which one of the
spin-split bands is completely occupied or completely empty as in magnetite 
\cite{Cullen}. Also this model can be viewed as a quarter-filled extended
Hubbard model in the large $U$ limit. Then double occupancy of a site is
forbidden and the nearest-neighbour Coulomb repulsion $V$ plays a crucial
role. Because for (spinless) half filling every second site is unoccupied on
the average one would naively expect a metallic state. Our main goal in this
work is to show that inter-site correlations $V$ can lead to a MI transition
even for a case with less than one electron per site. We are using two
different methods to study this model.

Firstly we employ a single site approach within a Hartree-Fock as well as
Hubbard-I type approximations (Sect.~\ref{1-site}). By using the
'Hubbbard-I'-type decoupling scheme which includes the effect of
nearest-neighbor charge fluctuations, we find that with increasing value of $V$ 
first a metal- insulator transition occurs with a gap in the excitation
spectrum, while at even larger values of V charge ordering appears. This is
opposite to the result in the simple Hartree-Fock approximation, i.e.
inter-site correlations favor the MI transition and suppress the CO. This
observation may indicate that CO is indeed not present for a rigid
checkerboard or pyrochlore lattice for any V/t ratio (it certainly is not in
the limit V/t$\rightarrow \infty$). Indeed in the compounds $AlV_2O_4$ and $LiV_2O_4$ 
(under pressure) where CO has been found it is accompanied by a
lattice distortion.

Secondly we start from a cluster representation of the model where the
intra-cluster correlations are taken into account exactly and the
inter-cluster terms are treated in Hartree Fock approximation (Sect.~\ref{cluster}). 
This transformation leads to an effective multi-orbital extended
Hubbard model. Again we find a M-I transition at a value similar to the
first approach. In view of the suggestion above we do not consider the
possibility of CO in this case, although within the Hartree Fock
approximation for the inter-cluster interactions it would presumably be
present in the ground state.

Furthermore, we investigated the evolution of quasiparticle bands in the 
various phases in the single-site and cluster approaches. Following the 
selfconsistently determined chemical potential and the formation of 
interband gaps allows to determine the critical value for the MI transition. 

\section{Single site approximation} \label{1-site}

The Hamiltonian for the spinless fermion model is given by \cite{Fulde}
\begin{equation}
H=-t\sum_{\left\langle ij\right\rangle } c_i^{\dagger }c_j
+V/2~\sum_{\left\langle ij\right\rangle }n_in_j  \label{1}
\end{equation}
where $V$ is the Coulomb repulsion between nearest neighbors denoted by $\left\langle
ij\right\rangle $. We refer to this spinless
Hamiltonian as the $t-V$ model.

First, let us briefly review previous work on the $t-V$ model at
half-filling and the closely related quarter-filled extended Hubbard model in the large $U$
limit for different lattices. In one dimension, the $t-V$ model can be
mapped onto an anisotropic Heisenberg model\cite{Ovchinnikov} and solved
exactly via Bethe ansatz. In this case, one finds a gapless metallic phase
for $V<2t$ and a gapped charge-ordered state for larger values of $V$. For $t-V$ 
model on a two-leg ladder, it was found from renomalization group
calculations that contrary to a single chain, the ladder becomes a Mott
insulator for arbitrarily small repulsive interactions $V$\cite{Donohue}. By
using the density matrix renormalization group method, Vojta\cite{Vojta} et.
al. studied the extended Hubbard model for two-leg ladders in the large $U$
limit. They found that the charge-ordered phase vanishes for $V<2.5t$ but
claimed that there will be a charge gap for all values of $V/t$. For a square
lattice, McKenzie\cite{McKenzie} et al. argued from slave-boson theory that
the insulating phase with charge order is destroyed below a critical value $V$ 
of order $t$ and the system becomes metallic.

Here we will study the $t-V$ model on the checkerboard lattice at
half-filling (one electron per two sites) with the aim to find out about
possible MI and CO transitions in this frustrated lattice which is a 2D
model for the pyrochlore lattice. At first we will employ the Green's
function method and various decoupling schemes within the single-site
representation. There are two sites per unit cell. For convenience, we
rewrite the Hamiltonian (1) in the following form
\begin{equation}
H=H_{t}+H_{V,MF}+H_{V}^{\prime }~~~,  \label{2}
\end{equation}
where  
\begin{equation}
H_{t}=-t\sum_{\mathbf{l}}[c_{\mathbf{l},2}^{\dagger }\left( c_{\mathbf{l}-
\mathbf{x}-\mathbf{y},1}+c_{\mathbf{l}-\mathbf{y},1}+c_{\mathbf{l}+\mathbf{x}
,2}\right) +c_{\mathbf{l},1}^{\dagger }\left( c_{\mathbf{l},2}+c_{\mathbf{l}+
\mathbf{x},2}+c_{\mathbf{l}-\mathbf{y},1}\right) +h.c.],  \label{3}
\end{equation}
\begin{eqnarray}
H_{V,MF} &=&V\sum_{\mathbf{l}}[n_{\mathbf{l},2}\left( \left\langle n_{
\mathbf{l}-\mathbf{x}-\mathbf{y},1}+n_{\mathbf{l}-\mathbf{y},1}+n_{\mathbf{l}
+\mathbf{x},2}\right\rangle \right) +\left\langle n_{\mathbf{l}
,2}\right\rangle \left( n_{\mathbf{l}-\mathbf{x}-\mathbf{y},1}+n_{\mathbf{l}-
\mathbf{y},1}+n_{\mathbf{l}+\mathbf{x},2}\right)  \nonumber \\
&&+n_{\mathbf{l},1}\left( \left\langle n_{\mathbf{l},2}+n_{\mathbf{l}+
\mathbf{x},2}+n_{\mathbf{l}-\mathbf{y},1}\right\rangle \right) +\left\langle
n_{\mathbf{l},1}\right\rangle \left( n_{\mathbf{l},2}+n_{\mathbf{l}+\mathbf{x
},2}+n_{\mathbf{l}-\mathbf{y},1}\right)  \nonumber \\
&&-\left\langle n_{\mathbf{l},2}\right\rangle \left( \left\langle n_{\mathbf{
l}-\mathbf{x}-\mathbf{y},1}+n_{\mathbf{l}-\mathbf{y},1}+n_{\mathbf{l}+
\mathbf{x},2}\right\rangle \right) -\left\langle n_{\mathbf{l}
,1}\right\rangle \left( \left\langle n_{\mathbf{l},2}+n_{\mathbf{l}+\mathbf{x
},2}+n_{\mathbf{l}-\mathbf{y},1}\right\rangle \right) ], \label{4}
\end{eqnarray}
\begin{equation}
H_{V}^{\prime }=V\sum_{\mathbf{l}}[\delta n_{\mathbf{l},2}\left( \delta n_{
\mathbf{l}-\mathbf{x}-\mathbf{y},1}+\delta n_{\mathbf{l}-\mathbf{y}
,1}+\delta n_{\mathbf{l}+\mathbf{x},2}\right) +\delta n_{\mathbf{l},1}\left(
\delta n_{\mathbf{l},2}+\delta n_{\mathbf{l}+\mathbf{x},2}+\delta n_{\mathbf{
l}-\mathbf{y},1}\right) ].  \label{5}
\end{equation}
Here we have introduced a charge fluctuation operator $\delta n_{\mathbf{l}
,i}=n_{\mathbf{l},i}-\left\langle n_{\mathbf{l},i}\right\rangle $ on sites 
$i=1,2$ of the $\mathbf{l}$-th cell. $H_{t}$ is the kinetic energy
term, $H_{V,MF}$ is mean-field part of the interaction term while $H_{V}^{\prime }$ is
the residual interaction part. In the following we only consider the
simplest charge ordered pattern, namely a staggered checkerboard pattern
with wave vector {$\mathbf{Q}$}$=\left( 0,0\right) $ illustrated in the
inset of Fig. 1. The average charge density on different sites in the cell is 
$\left\langle n_{\mathbf{l},i}\right\rangle =\frac{1}{2}-\left( -1\right)
^{i}n$, where the order parameter $n$ means a charge disproportionation within
each unit cell.

\begin{figure}[tbp]
\includegraphics[clip=true,width=80mm]{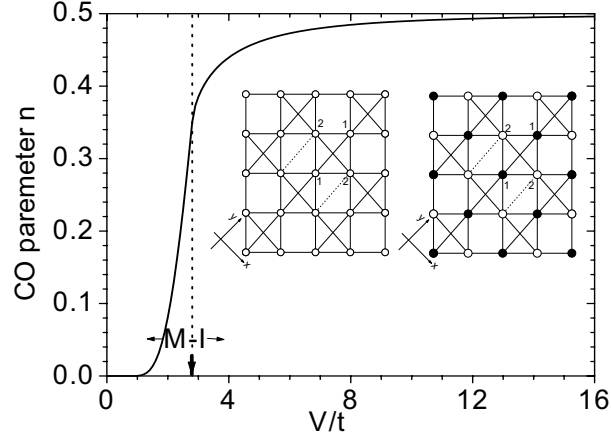}
\caption{Charge order parameter $n$ as a function of $V/t$ within a
mean-field approximation. Dotted line indicates the critical point of the
metal-insulator transition. Inset is an illustration of the checkerboard
lattice with (right) and without (left) charge order. The $\mathbf{l}$-th unit cell 
is indicated by dotted lines. The wave vector of the staggered CO is {$\mathbf{Q}$}
$=\left( 0 ,0 \right) $ due to the two sites per unit cell. Here charge order 
occurs before the metal-insulator transition takes place.}
\label{fig:Fig1}
\end{figure}

The electron propagation is described by a retarded Green's function (for
simplicity the conventional symbol R is omitted) 
\begin{equation}
G_{i,j}\left( \mathbf{l}-\mathbf{l}^{\prime },\omega \right) =\left\langle
\left\langle c_{\mathbf{l}i}|c_{\mathbf{l}^{\prime }j}^{\dagger
}\right\rangle \right\rangle _{\omega }=\int \left\langle \left\langle c_{
\mathbf{l}i}(t)|c_{\mathbf{l}^{\prime }j}^{\dagger }\right\rangle
\right\rangle e^{i\omega t}dt.~~~  \label{6}
\end{equation}
where 
\begin{equation}
\left\langle \left\langle c_{\mathbf{l}i}(t)|c_{\mathbf{l}^{\prime
}j}^{\dagger }\right\rangle \right\rangle =-i\theta \left( t\right)
\left\langle \{ c_{\mathbf{l}i}\left( t\right) ,c_{\mathbf{l}^{\prime
}j}^{\dagger }\left( 0\right) \}\right\rangle .  \label{7}
\end{equation}
and $\{,\}$ denotes the anticommutator. The above ($2\times 2$)-matrix Green's function must satisfy the equation 
\begin{equation}
\omega \left\langle \left\langle A|B\right\rangle \right\rangle _{\omega
}=\left\langle {\ }\{ {A,B}\} \right\rangle +\left\langle
\left\langle \left[ A,H\right] |B\right\rangle \right\rangle _{\omega }.
\label{8}
\end{equation}
By introducing the Fourier transformation 
\begin{equation}
G_{i,j}\left( \mathbf{{k},\omega }\right) =\frac{1}{N}\sum_{\mathbf{k}}e^{i
\mathbf{{k}\cdot \left( R_{l}-R_{l}^{\prime }\right) }}G_{i,j}\left( \mathbf{
l}-\mathbf{l}^{\prime },\omega \right) ,  \label{9}
\end{equation}
the equation (8) can be now written explicitly as 
\begin{equation}
\left( \omega \widehat{1}+\widehat{\Lambda }\left( \mathbf{k}\right) \right) 
\widehat{G}\left( \mathbf{{k},\omega }\right) =\widehat{1}+V\widehat{\Gamma }
\left( \mathbf{{k},\omega }\right) .  \label{10}
\end{equation}
with
\begin{equation}
\widehat{\Lambda }\left(\mathbf{k}\right)=\left( 
\begin{array}{ll}
-\left( 3-2n\right) V+2t\cos k_{y}~~~ & 4te^{-i\frac{k_{x}+k_{y}}{2}}\cos 
\frac{k_{x}}{2}\cos \frac{k_{y}}{2} \\ 
4te^{i\frac{k_{x}+k_{y}}{2}}\cos \frac{k_{x}}{2}\cos \frac{k_{y}}{2}~~~ & 
-\left( 3+2n\right) V+2t\cos k_{x}
\end{array}
\right) ,  \label{11}
\end{equation}
where $\widehat{1}$ is the unit matrix and $\widehat{{\Gamma }}\left( \mathbf{{
k},\omega }\right) $ is the Fourier transformation of the ($2\times 2$
)-matrix $\Gamma _{i,j}\left( \mathbf{l}-\mathbf{l}^{\prime }, \omega \right) $
($i,j=1,2$) which is defined as
\begin{equation}
\Gamma _{i,j}\left( \mathbf{l}-\mathbf{l}^{\prime },\omega \right)
=\left\langle \left\langle c_{\mathbf{l},i}\sum_{\left( n.n.\right) \in
\left( \mathbf{l},i\right) }\delta n_{\left( n.n.\right) }|c_{\mathbf{l}
^{\prime },j}^{\dagger }\right\rangle \right\rangle _{\omega }.  \label{12}
\end{equation}
Here the summation is over six nearest-neighbor sites (n.n.) surrounding the
site ($\mathbf{l},i$)
\begin{figure}[tbp]
\includegraphics[clip=true,width=100mm]{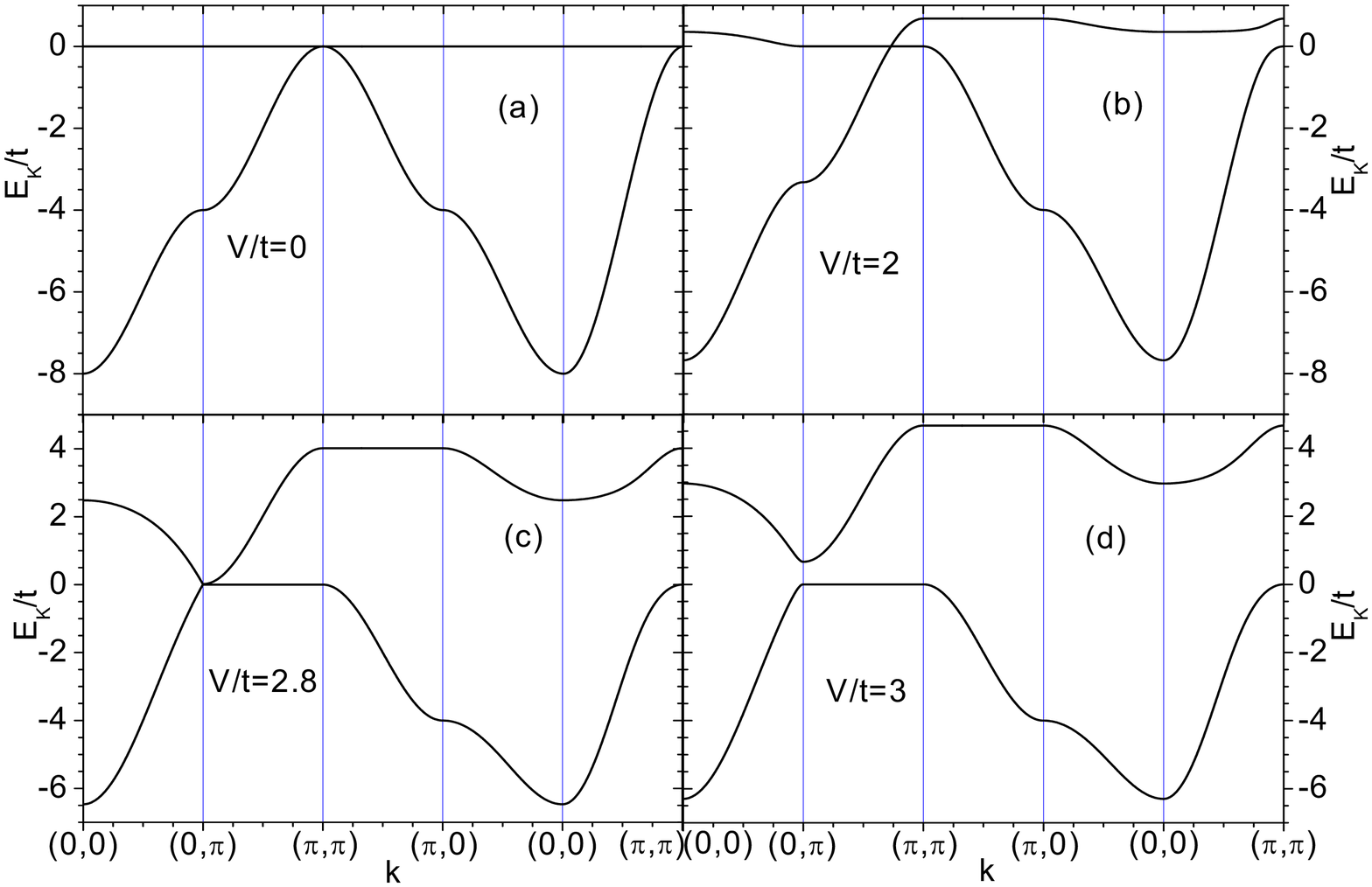}
\caption{Quasiparticle dispersion within mean-field theory for different
value of $V/t$. The chemical potential $\protect\mu $ is fixed at zero energy. 
(a) is the noninteracting case without charge order. These two bands touch 
at $\left( \protect\pi,\protect\pi \right) $ and the upper band is flat. (b) 
is in the metallic phase with small charge order. The two bands still touch 
while the upper band become dispersive due to the inequivalence of diagonal 
hopping. (c) is exactly at the critical point. The two bands touch at 
$\left( 0,\protect\pi \right) $. (d) is in the insulating phase. 
The two bands seperate.}
\end{figure}

\subsection{Mean-field approximation}

In the first step we may decouple 
\begin{equation}
\Gamma _{i,j}\left( \mathbf{l}-\mathbf{l}^{\prime },\omega \right) \approx
\left\langle \sum_{\left( n.n.\right) \in \left( \mathbf{l},i\right) }\delta
n_{\left( n.n.\right) }\right\rangle \left\langle \left\langle c_{\mathbf{l}
,i}|c_{\mathbf{l}^{\prime },j}^{\dagger }\right\rangle \right\rangle
_{\omega }=0,  \label{13}
\end{equation}
which leads to a mean-field approximation and the quasiparticle dispersions
are given by 
\begin{eqnarray}
E_{\mathbf{k}}^{\pm }/t &=&-(\cos k_{x}+\cos k_{y}-3V/t)  \nonumber \\
&&\pm \lbrack 4(1+\cos k_{x})(1+\cos k_{y})+(2nV/t-\cos k_{x}+\cos
k_{y})^{2}]^{1/2}.  \label{14}
\end{eqnarray}
Now the intra-cell charge disproportionation $n$ has to be determined from
the following self-consistent equation 
\begin{equation}
n=\frac{1}{2N}\sum_{\mathbf{k}}\frac{\left( 2nV/t-\cos k_{x}+\cos
k_{y}\right) }{[4(1+\cos k_{x})(1+\cos k_{y})+(2nV/t-\cos k_{x}+\cos
k_{y})^{2}]^{1/2}}.  \label{15}
\end{equation}
The calculated phase diagram is shown in Fig. 1. In the present mean-field
approximation ($\widehat{\Gamma }\equiv 0$) the metallic phase is already
charge ordered before a charge-transfer-type metal-insulator
phase transition takes place at $V=2.8t$. In the non-interacting case $V=0$ (Fig. (2a)), 
the upper band is flat while the lower band is dispersive. It
touches the flat band at $\left( \pi ,\pi \right) $. As shown in Fig. (2b, 2c) with 
increasing intersite repulsion $V$ the flat band becomes
increasingly dispersive with increasing charge order and the touching point
moves from $\left( \pi ,\pi \right) $ towards $\left( 0,\pi \right) .$ The
dispersion of the previously flat band is due to the inequivalence of
diagonal hopping (see inset of Fig. 1) induced by charge order. Finally, for $V>2.8t$ the 
two bands separate as shown in Fig. (2d). In the half-filled case, the
lower band is fully occupied and the upper band is empty resulting in a
charge-transfer-type insulator.

\subsection{'Hubbard I'- approximation for the inter-site correlations}

The neglect of correlations overestimates the tendency to CO symmetry breaking. 
Therefore, in the strongly correlated case ($V\gg t$), the mean-field
results are unreliable and it is necessary to consider equations of motion
of higher-order Green's functions $\Gamma _{i,j}\left( \mathbf{l}-
\mathbf{l}^{\prime },\omega \right) $ which can be written as 
\begin{eqnarray}
&&\left( \omega -V \sum_{\left( n.n.\right) \in
\left( \mathbf{l},i\right) }\left\langle n_{\left( n.n.\right)}\right\rangle\right)
\left\langle \left\langle c_{\mathbf{l},i}\sum_{\left( n.n.\right) \in
\left( \mathbf{l},i\right) }\delta n_{\left( n.n.\right) }|c_{\mathbf{l}
^{\prime },j}^{\dagger }\right\rangle \right\rangle _{\omega }  \nonumber \\
&=&\sum_{\left( n.n.\right) \in \left( \mathbf{l},i\right) } \delta
_{\left( \mathbf{l}^{\prime },j \right),\left( n.n.\right)} 
\left\langle c_{\mathbf{l}i}c_{\left(n.n.\right)}^{\dagger} \right\rangle 
-t\sum_{\left( n.n.\right) \in \left( \mathbf{l},i\right) }\left\langle \left\langle 
c_{\left( n.n.\right) }\delta n_{\mathbf{l},i}|c_{\mathbf{l}^{\prime },j}^{\dagger }
\right\rangle\right\rangle _{\omega }      \nonumber \\
&&-t\sum_{\left( n.n.\right)
\in \left( \mathbf{l},i\right) } \left(\left\langle n_{\mathbf{l},i}\right\rangle 
- \left\langle n_{n.n.}\right\rangle \right) \left\langle \left\langle
c_{\left( n.n.\right) }|c_{\mathbf{l}^{\prime },j}^{\dagger }\right\rangle
\right\rangle _{\omega }  \nonumber \\
&&-t\sum_{\left( n.n.\right) \in \left( \mathbf{l},i\right) }\sum_{\left(
n.n.\right) ^{\prime }\in \left( \mathbf{l},i\right) }\left( 1-\delta
_{\left( n.n.\right) ^{\prime },\left( n.n.\right) }\right) \left\langle
\left\langle c_{\left( n.n.\right) ^{\prime }}\delta n_{\left( n.n.\right)
}|c_{\mathbf{l}^{\prime },j}^{\dagger }\right\rangle \right\rangle _{\omega }
\nonumber \\
&&-t\sum_{\left( n.n.\right) \in \left( \mathbf{l},i\right) }\sum_{\left(
n.n.\right) ^{\prime }\in \left( n.n.\right) }\left( 1-\delta _{\left(
n.n.\right) ^{\prime },\left( \mathbf{l},i\right) }\right) \left\langle
\left\langle c_{\mathbf{l},i}c_{\left( n.n.\right) }^{+}c_{\left(
n.n.\right) ^{\prime }}|c_{\mathbf{l}^{\prime },j}^{\dagger }\right\rangle
\right\rangle _{\omega }  \nonumber \\
&&+t\sum_{\left( n.n.\right) \in \left( \mathbf{l},i\right) }\sum_{\left(
n.n.\right) ^{\prime }\in \left( n.n.\right) }\left( 1-\delta _{\left(
n.n.\right) ^{\prime },\left( \mathbf{l},i\right) }\right) \left\langle
\left\langle c_{\mathbf{l},i}c_{\left( n.n.\right) ^{\prime }}^{+}c_{\left(
n.n.\right) }|c_{\mathbf{l}^{\prime },j}^{\dagger }\right\rangle
\right\rangle _{\omega }  \nonumber \\
&&+V\left\langle \left\langle c_{\mathbf{l},i}\left( \sum_{\left(
n.n.\right) \in \left( \mathbf{l},i\right) }\delta n_{\left( n.n.\right)
}\right) ^{2}|c_{\mathbf{l}^{\prime },j}^{\dagger }\right\rangle
\right\rangle _{\omega }, \label{16}
\end{eqnarray}
Here we adopt the following approximations, which consist in an extension of the on-site
Hubbard I decoupling scheme\cite{Hubbard} to spinless fermion on a
checkerboard lattice with intersite Coulomb interaction: 
\begin{eqnarray}
&&\left\langle \left\langle c_{\mathbf{l},i}\left( c_{\left( n.n.\right)
}^{+}c_{\left( n.n.\right) ^{\prime }}-c_{\left( n.n.\right) ^{\prime
}}^{+}c_{\left( n.n.\right) }\right) |c_{\mathbf{l}^{\prime },j}^{\dagger
}\right\rangle \right\rangle _{\omega }  \nonumber \\
&\simeq &\left( \left\langle c_{\left( n.n.\right) }^{+}c_{\left(
n.n.\right) ^{\prime }}\right\rangle -\left\langle c_{\left( n.n.\right)
^{\prime }}^{+}c_{\left( n.n.\right) }\right\rangle \right) \left\langle
\left\langle c_{\mathbf{l},i}|c_{\mathbf{l}^{\prime },j}^{\dagger
}\right\rangle \right\rangle _{\omega }=0. \label{17}
\end{eqnarray}
In equation (\ref{16}), the fourth term involves both the nearest- and
next-nearest-neighbor charge correlations and we neglect the latter. When
treating the last term in equation (\ref{16}) we also neglect the more distant charge
correlations and approximate as
\begin{figure}[tbp]
\includegraphics[clip=true,width=60mm]{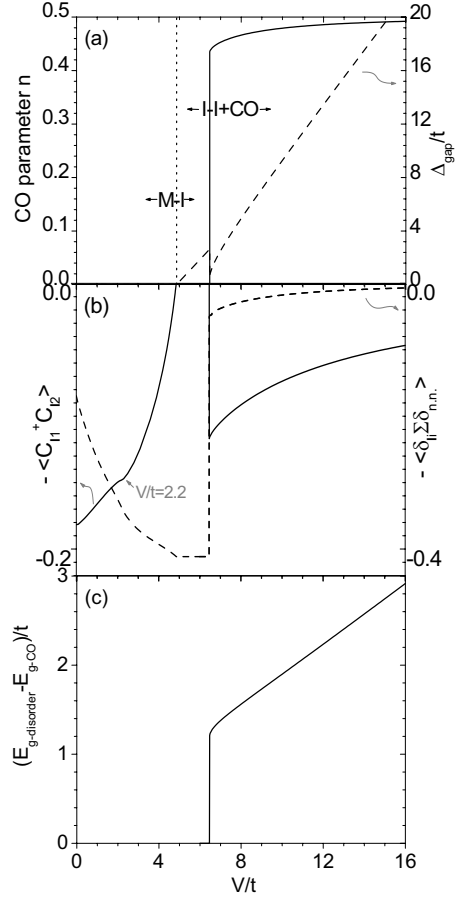}
\caption{Results within Hubbard I-like approximation. (a) charge order (CO)
parameter $n$ versus $V/t$. The dotted line indicates the phase boundary
between the metallic and insulating phase without charge ordering. Contrast
to the mean-field approximation, the charge-ordered phase transition is of
first order and occurs after a Mott-Hubbard-like metal-insulator transition
takes place. The dashed line shows the quasiparticle gap versus $V/t$ in
both insulating phases. (b) Hopping amplitude (solid line) from site 2 to site 1 
($\left\langle c_{\mathbf{l}1}^{\dagger }c_{\mathbf{l}2}\right\rangle $) and 
nearest-neighbor charge fluctuation correlations $\delta n_{\mathbf{l}i}\sum_{\left( n.n.\right) 
\in \left( \mathbf{l},i\right) }\delta n_{\left( n.n.\right) }$ (dashed line) versus 
$V/t$. In the insulating phase without charge order, the expectation value 
$\left\langle c_{\mathbf{l}1}^{\dagger }c_{\mathbf{l}2}\right\rangle $ is equal
to 0 while $\delta n_{\mathbf{l}i}\sum_{\left( n.n.\right) \in \left( \mathbf{l},i\right) }
\delta n_{\left( n.n.\right) }$ keeps constant and maximum. Note that $\delta n_{\mathbf{l}i}
\sum_{\left( n.n.\right) \in \left( \mathbf{l},i\right) }\delta n_{\left( n.n.\right) }$ 
is site-independent for all value of $V/t$. (c) Energy difference between ordered 
and disordered state versus $V/t$. For $V>V_{c2}$, the charge-ordered state has lower energy.}
\end{figure}
\begin{eqnarray}
&&\left\langle \left\langle c_{\mathbf{l},1}\left( \sum_{\left( n.n.\right)
\in \left( \mathbf{l},1\right) }\delta n_{\left( n.n.\right) }\right)
^{2}|c_{\mathbf{l}^{\prime },j}^{\dagger }\right\rangle \right\rangle
_{\omega }  \nonumber \\
&\simeq &\left\langle \left( \sum_{\left( n.n.\right) \in \left( \mathbf{l}
,1\right) }\delta n_{\left( n.n.\right) }\right) ^{2}\right\rangle
\left\langle \left\langle c_{\mathbf{l},1}|c_{\mathbf{l}^{\prime
},j}^{\dagger }\right\rangle \right\rangle _{\omega }  \nonumber \\
&\simeq &\left( \frac{3}{2}-6n^{2}+2\left\langle \delta n_{\mathbf{l}
,2}\sum_{\left( n.n.\right) \in \left( \mathbf{l},2\right) }\delta n_{\left(
n.n.\right) }\right\rangle \right) \left\langle \left\langle c_{\mathbf{l}
,1}|c_{\mathbf{l}^{\prime },j}^{\dagger }\right\rangle \right\rangle
_{\omega }.   \label{18}
\end{eqnarray}
Finally the equation of motion for the higher-order Green's function $\Gamma
_{i,j}\left( \mathbf{{k},\omega }\right) $ can be written as
\begin{equation}
\left( \omega \widehat{1}+\widehat{\Lambda }\left( \mathbf{k}\right) \right) 
\widehat{\Gamma }\left( \mathbf{{k},\omega }\right) =\widehat{B}\left(\mathbf{k}\right)+
\widehat{M}\left(\mathbf{k}\right)\widehat{G}\left( \mathbf{{k},\omega }\right) .
\label{19}
\end{equation}
where
\begin{equation}
\widehat{M}\left(\mathbf{k}\right)=\left( 
\begin{array}{ll}
~~~~~~~~~~~~~VK_{2}~~~~~~~ & -8nte^{-i\frac{k_{x}+k_{y}}{2}}\cos \frac{k_{x}
}{2}\cos \frac{k_{y}}{2} \\ 
8nte^{i\frac{k_{x}+k_{y}}{2}}\cos \frac{k_{x}}{2}\cos \frac{k_{y}}{2}~~~ & 
~~~~~~~~~~~~~VK_{1}~~~~~~~
\end{array}
\right) , \label{20A}
\end{equation}
and 
\begin{equation}
\widehat{B}\left(\mathbf{k}\right)=\left( 
\begin{array}{cc}
2\left\langle c_{\mathbf{l}1}c_{\mathbf{l-y}1}^{\dagger }\right\rangle\cos k_{y} & 4\left\langle
c_{\mathbf{l}1}c_{\mathbf{l}2}^{\dagger }\right\rangle e^{-i\frac{k_{x}+k_{y}}{2}}\cos \frac{k_{x}}{2}\cos 
\frac{k_{y}}{2} \\ 
4\left\langle c_{\mathbf{l}1}c_{\mathbf{l}2}^{\dagger }\right\rangle e^{i\frac{k_{x}+k_{y}}{2}}\cos \frac{k_{x}}{2}\cos \frac{k_{y}}{2} & 
2\left\langle c_{\mathbf{l}2}c_{\mathbf{l+x}2}^{\dagger }\right\rangle\cos k_{x}
\end{array}
\right) .  \label{20B}
\end{equation}
Here we used the definition 
\begin{equation}
K_{i}=\left( \frac{3}{2}-6n^{2}+2\left\langle \delta n_{\mathbf{l}
,i}\sum_{\left( n.n.\right) \in \left( \mathbf{l},i\right) }\delta
n_{\left( n.n.\right)}\right\rangle \right) .  \label{21}
\end{equation}
By solving equation (\ref{19}) with respect to $\widehat{\Gamma }$ and after substituting 
$\widehat{\Gamma }$ into equation (\ref{10}) one obtains the final solution for the
Green's function $\widehat{G}$. For the electron concentration $1/2$ (one
electron per unit cell), the chemical potential $\mu $, the charge
disproportionation $n$ and the hopping amplitute $\left\langle c_{\mathbf{l}1}c_{\mathbf{l}2}^{\dagger }\right\rangle$ 
are calculated from the following set of self-consistent equations: 
\begin{equation}
1=\frac{1}{N/2}\sum_{\bf{k}}\int_{-\infty }^{\mu }d\omega \left( -\frac{1
}{\pi }\right) \mathop{\rm Im}\left[ G_{11}\left( \bf{k},\omega \right)
+G_{22}\left( \bf{k},\omega \right) \right] ,  \label{22}
\end{equation}
\begin{equation}
n=\frac{1}{N/2}\sum_{\bf{k}}\int_{-\infty }^{\mu }d\omega \left( -\frac{1
}{\pi }\right) \mathop{\rm Im}\left[ \frac{G_{11}\left( \bf{k},\omega \right)
-G_{22}\left( \bf{k},\omega \right) }{2}\right] ,  \label{23}
\end{equation}
\begin{equation}
\left\langle c_{\mathbf{l}1}c_{\mathbf{l}2}^{\dagger }\right\rangle=\frac{1}{N/2}\sum_{\bf{k}}\int_{-\infty }^{\mu }d\omega \left( -\frac{1
}{\pi }\right) \mathop{\rm Im}G_{12}(\bf{k},\omega ).  \label{24}
\end{equation}
The higher-order Green's function $\Gamma _{i,j}\left( \mathbf{{k},\omega }\right) $ 
can be easily derived from equation (\ref{10}). Therefore the nearest-neighbor charge fluctuation 
correlation $\left\langle \delta n_{\mathbf{l},i}\sum_{\left( n.n.\right) \in \left( \mathbf{l},i\right) }
\delta n_{\left( n.n.\right)}\right\rangle$ can be calculated as 
\begin{equation}
\left\langle \delta n_{\mathbf{l}i}\sum_{\left( n.n.\right) \in
\left( \mathbf{l},i\right) }\delta n_{\left( n.n.\right) }
\right\rangle =\frac 1{N/2}\sum_{\mathbf{k}}\int_{-\infty }^\mu d\omega
\left( -\frac 1\pi \right) \mathop{\rm Im}\Gamma _{i,i}\left( \mathbf{{k}
,\omega }\right) .  \label{24b}
\end{equation}
Furthermore the hopping amplitude $\left\langle c_{\mathbf{l}1}c_{\mathbf{l-y}1}^{\dagger }\right\rangle$ 
and $\left\langle c_{\mathbf{l}2}c_{\mathbf{l+x}2}^{\dagger }\right\rangle$ can be 
also determined from $G_{11}\left( \bf{k},\omega \right)$ and $G_{22}\left( \bf{k},\omega \right)$ respectively. 

The one-particle retarded Green's function exhibits four poles given by the roots of 
$D=0$ where $D$ is defined as 
\begin{eqnarray}
&&D=\left( \omega -E_{\mathbf{k}}^{+}\right) ^{2}\left( \omega -E_{\mathbf{k}
}^{-}\right) ^{2}+V^{4}K_{1}K_{2}+16n^{2}V^{2}t^{2}\left( 1+\cos
k_{x}\right) \left( 1+\cos k_{y}\right)   \nonumber \\
&&-V^{2}K_{1}[\left( \omega -\left( 3-2n\right) V+2t\cos k_{y}\right)
^{2}-4t^{2}\left( 1+\cos k_{x}\right) \left( 1+\cos k_{y}\right) ]  \nonumber \\
&&-V^{2}K_{2}[\left( \omega -\left( 3+2n\right) V+2t\cos k_{x}\right)
^{2}-4t^{2}\left( 1+\cos k_{x}\right) \left( 1+\cos k_{y}\right) ]
\label{25}
\end{eqnarray}
As a consequence, four bands are obtained. For $V=0$, two branches of momentum-dependent
dispersions are obtained from equation (\ref{25}): $E_{\mathbf{{k}}
}^{-}|_{V=0}=-2t(\cos k_{x}+\cos k_{y}+1)$ and $E_{\mathbf{{k}}}^{+}|_{V=0}=2t$ as
expected. When $t=0$, two poles $\omega =2V$ and $\omega =4V$ are obtained
from above equations, which describes one hole or one particle excitation.
\begin{figure}[tbp]
\includegraphics[clip=true,width=100mm]{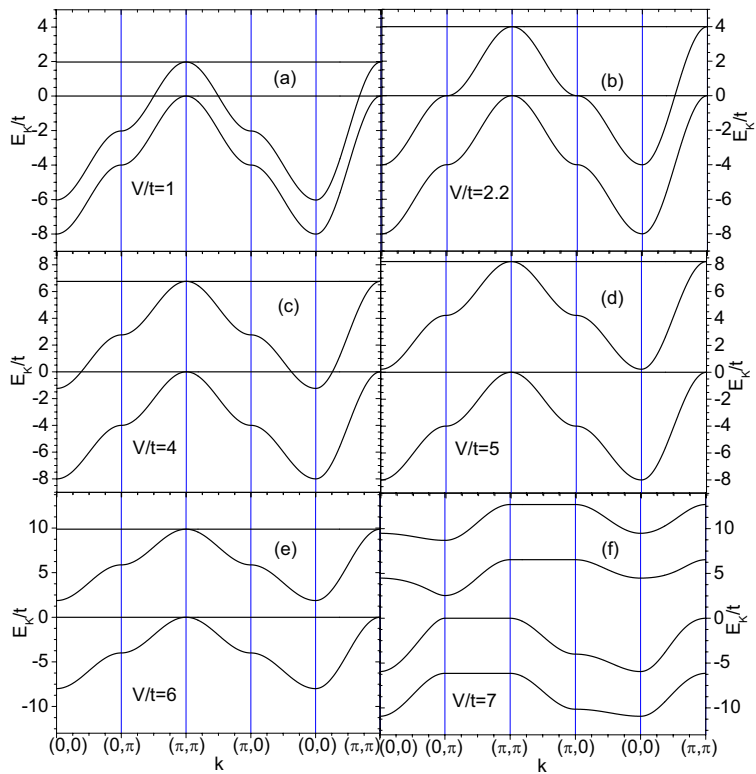}
\caption{evolution of the quasiparticle spectrum within a
Hubbard I-like approximation for different value of $V/t$. The chemical potential $\protect\mu $ 
is fixed at zero energy. $V/t=0$ is already shown in Fig. 2 (a). Figs. (a), (b) and (c) are in 
the metallic phase ($0<V<V_{c1}=4.86t$). Figs. (d) and (e) are in the insulating phase without 
CO symmetry breaking ($V_{c1}<V<V_{c2}=6.47t$). Reminiscent of the upper and lower 
Hubbard band splitting, the metal-insulator phase transition at $V_{c1}$ is of Mott-Hubbard type. 
Fig. (f) is in the insulating phase with CO ($V>V_{c2}$). }
\end{figure}

In Figure (3a) we show the phase diagram within the previously used
decoupling scheme. First, a metal-insulator transition occurs at $
V_{c1}=4.86t$ and then charge ordering appears at $V_{c2}=6.47t$. In the
interval $V>V_{c1}$ and $V<V_{c2}$ a gap opens and increases with increasing
value of $V$. At $V_{c2}$, it drops to a smaller value. For $V>V_{c2}$, the
gap increases again monotonously. In the charge ordered phase, there exist
always two self-consistent solutions, i.e., a charge ordered and a
disordered one. By comparing the energy of these two states, (see Fig. 3 (c)),
we find for $V>V_{c2}$ that the charge ordered state has lower energy. The
phase transition is of first order. Fig. 3 (b) shows that at the
metal-insulator transition the hopping amplitude 
$\left\langle c_{\mathbf{l}1}c_{\mathbf{l}2}^{\dagger }\right\rangle$ vanishes.

In the following we want to explain this feature in more detail. For that
purpose we study the poles of the retarded Green's function as a function 
of increasing value of $V/t$. Consider first the trivial case of $V=0$ 
(see Fig. 2 (a)). The states in the dispersive band are of bonding and in 
the flat band of antibonding character. As seen in Fig. 4 (a), (b) and (c) these bands split 
into four when $0<V<V_{c1}$ and the Hubbard-I like approximation is made. 
The retarded Green's function has therefore four poles for each $\mathbf{k}$ vector. 
In the regime $0<V<V_{c1}$,
states in the antibonding flat band become more and more occupied as $V$
increases. Therefore the expectation value $\left\langle c_{\mathbf{l}
1}^{\dagger }c_{\mathbf{l}2}\right\rangle $ decreases since an equal
occupational probability of a bonding and antibonding state implies that $
\left\langle c_{\mathbf{l}1}^{\dagger }c_{\mathbf{l}2}\right\rangle =0$.
This is the case in the regime $V_{c1}<V<V_{c2}$ shown in Figs. 4 (d), (e).
Note that a correlation gap has opened above the two occupied lower bands 
which implies a Mott-Hubbard type insulating state and the nearest-neighbor 
charge fluctuation correlations $\delta n_{\mathbf{l}i}\sum_{\left( n.n.\right) 
\in \left( \mathbf{l},i\right) }\delta n_{\left( n.n.\right) }$ remain constant. 
When $V>V_{c2}$, then due to symmetry breaking one can no longer distinguish between
bonding and antibonding state and $\left\langle c_{\mathbf{l}1}^{\dagger }c_{
\mathbf{l}2}\right\rangle \neq 0$ (see Figs. 4 (f)). As $V$ continues to
increase, charge order become more and more pronounced and $\left\langle c_{
\mathbf{l}1}^{\dagger }c_{\mathbf{l}2}\right\rangle $ decreases again
monotonously. This explains the behavior of $\left\langle c_{\mathbf{l}
1}^{\dagger }c_{\mathbf{l}2}\right\rangle $ shown in Fig. 3. The inflexion point 
at $V/t=2.2$ shown in Fig. 3 (b) is due to the lower flat band crossing the 
inflexion pionts (maximum of DOS) of the upper dispersive band shown in Fig. 4 (b). 
For $V>V_{c2}$, CO strongly suppresses the nearest-neighbor charge fluctuation 
correlations $\delta n_{\mathbf{l}i}\sum_{\left( n.n.\right) \in \left( \mathbf{l},i\right) }\delta n_{\left( n.n.\right) }$ 
which is site-independent for all value of $V/t$ as expected. 

In this section we have studied the half-filled (one electron per two
sites) spinless $t-V$ model on a
checkerboard lattice within mean field theory and a 'Hubbard-I type'
approach. In the former approximation which overestimates the tendency
to symmetry breaking, CO appears before the system becomes an
insulator. By using a Green's function approach and a decoupling scheme 
that includes the effect of nearest-neighbor charge fluctuations, however, 
it was shown that \emph{first} a MI transition into an insulating state 
without CO takes place and only for larger V eventually CO appears. It is 
not entirely clear, however, how realistic the appearance of CO is. In the 
limit (V/t $\rightarrow\infty$) when the hopping vanishes, the ground state
for half filling is macroscopically degenerate \cite{Fulde,Anderson} and no 
CO is present \cite{note}. It is therefore possible that CO obtained here for the rigid
checkerboard lattice above $V_{c2}$ may be due to the employed Hubbard-I 
type approximation. In fact, the charge ordered state obtained here is 
one of the macroscopically degenerate states which must obey the tetrahedron 
rule (Anderson rule) \cite{Anderson}. If there would exist corresponding 
lattice distortion, for example, compressed along the diagonal direction \cite{Fujimoto}, 
the system would select such a CO state shown in the inset of Fig. 1 out 
of the macroscopically degenerate states.
Indeed in some 3D pyrochlore compounds which exhibit CO
as mentioned in the introduction, it is always accompanied by symmetry
lowering lattice distortions which remove the frustration by
introducing inequivalent bond lengths in the tetrahedrons of the
corner sharing lattice.

\section{Cluster approximation} \label{cluster}

We want to supplement the previous calculation based on a
Hubbard I type of approximation by another one where the strong electron
correlations are treated exactly within a cluster but within mean-field
approximation outside the cluster. Within this scheme we want to determine
the critical interaction $V_{c1}$ at which a gap opens in the excitation
spectrum when the case of half-filling is considered. We do not care here
about a possible charge order at large value of $V$ because, as pointed out
before, that may turn out to be an artefact of the involved
approximations. 

We divide the checkerboard lattice into sublattice A and B of the plaquette
so that each sublattice contains $N/4$ units where $N$ is the number of
sites. Fig. 5 shows black (basic) clusters linked accross white square. The
Hamiltonian (1), $H=H_{t}+H_{V}$, is decomposed into 
\begin{equation}
H_{t}=H_{t}^{(intra)}+H_{t}^{(inter)};~~~H_{V}=H_{V}^{(intra)}+H_{V}^{(inter)}
\label{27}
\end{equation}
Here each of the terms, $H_{t}^{(intra)}$ and $H_{V}^{(intra)}$, is a sum
over $l=1,...,N/4$ decoupled basic clusters while $H_{t}^{(inter)}$ and $
H_{V}^{(inter)}$are due to the intercluster coupling. Specifically the
decoupled clusters are treated by the Hamiltonian 
\begin{equation}
H_{t}^{(intra)}+H_{V}^{(intra)}=\sum_{\mathbf{l}}
H_{t-V}^{(intra)}\left( \mathbf{l}\right)   \label{28}
\end{equation}
\begin{equation}
H_{t-V}^{(intra)}\left( \mathbf{l}\right) =\sum_{i,j}M_{ij}\left( 0\right)
\left( -tc_{i\mathbf{l}}^{\dagger }c_{j\mathbf{l}}+\frac{V}{2}n_{i\mathbf{l}
}n_{j\mathbf{l}}\right) .  \label{29}
\end{equation}
Here $i,j=1,...,4$ denote the sites within a cluster with lattice vector $
\mathbf{l}$ and $M_{ij}\left( 0\right) $ is a ($4\times 4$)-matrix with
elements $M_{ij}\left( 0\right) =1$, if $i\neq j$, and $M_{ii}\left(
0\right) =0$. Furthermore $n_{i\mathbf{l}}=c_{i\mathbf{l}}^{\dagger }c_{i
\mathbf{l}}$. Thus, $H_{t-V}^{(intra)}\left( \mathbf{l}\right) $ represents
an intracluster Hamiltonian to be solved below.

To write down in a systematic way the intercluster coupling given by $
H_{t}^{(inter)}$ and $H_{V}^{(inter)}$, we refer to Fig. 5 where a fragment
of the checkerboard lattice is shown with the origin located at the center
of a basic cluster. There are eight neighboring basic clusters connected to
a given one with the translation vectors \{\mbox{\boldmath
$\tau$} \} $=$ \mbox{\boldmath $\tau$}$_{1}$, \mbox{\boldmath
$\tau$}$_{2}$,...,\mbox{\boldmath $\tau$}$_{8}$. Three of them are depicted
in Fig. 5. The intercluster electron hopping and Coulomb repulsion terms can
be written as  
\begin{equation}
H_{t}^{(inter)}=-t\sum_{\mathbf{l}}\sum_{\left\{ \mbox{\boldmath
$\tau$}\right\} }\sum_{i,j}M_{ij}\left( \mbox{\boldmath $\tau$}\right) c_{i,
\mathbf{l}+\mbox{\boldmath
$\tau$}}^{\dagger }c_{j\mathbf{l}}  \label{31}
\end{equation}
\begin{equation}
H_{V}^{(inter)}=\frac{V}{2}\sum_{\mathbf{l}}\sum_{\left\{ \mbox{\boldmath
$\tau$}\right\} }\sum_{i,j}M_{ij}\left( \mbox{\boldmath
$\tau$}\right) n_{i,\mathbf{l}+\mbox{\boldmath $\tau$ }}n_{j\mathbf{l}}
\label{32}
\end{equation}
where the eight ($4\times 4$)-matrices $M_{ij}\left( \mbox{\boldmath $\tau$ }
\right) $ are specified by referring to Fig. 5. Consider first the matrix $
M_{ij}\left( \mbox{\boldmath
$\tau$}_{1}\right) $ that connects two neighboring clusters by $
\mbox{\boldmath $\tau$}_{1}$-translation. According to Fig. 5, only two
individual two-site bonds contribute to this connection (reading from the
right to the left): ($ij$)=(14), (23). Therefore, we define $M_{14}\left( 
\mbox{\boldmath $\tau$}_{1}\right) =M_{23}\left( \mbox{\boldmath $\tau$}
_{1}\right) =1$ and $M_{ij}\left( \mbox{\boldmath $\tau$}_{1}\right) =0$
otherwise. In a similar way, one finds $M_{13}\left( \mbox{\boldmath $\tau$}
_{2}\right) =1,M_{12}\left( \mbox{\boldmath $\tau$}_{3}\right) =M_{43}\left( 
\mbox{\boldmath $\tau$}_{3}\right) =1$ and $M_{ij}\left( 
\mbox{\boldmath
$\tau$}_{2}\right) =M_{ij}\left( \mbox{\boldmath $\tau$}_{3}\right) =0$
otherwise. With this procedure, the other five matrices $M_{ij}\left( 
\mbox{\boldmath
$\tau$ }\right) $ can be easily found as well.

\begin{figure}[tbp]
\includegraphics[clip=true,width=30mm]{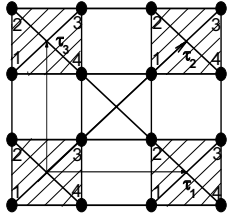}
\caption{Fragment of the checkerboard lattice. Only four neighboring basic
clusters connected by elementary translation vectors $\protect\tau_1$, 
$\protect\tau_2$ and $\protect\tau_3$ are shown. Within each basic cluster
the lattice sites are denoted by 1,...,4.}
\end{figure}

Now we solve the intra-cluster eigenvalue problem by diagonalizing the
Hamiltonian $H_{t-V}^{(intra)}\left( \mathbf{l}\right) $ from (\ref{29})
separately in each $n$-particle ($n$=0,1,...,4) sector 
\begin{equation}
H_{t-V}^{(intra)}(\mathbf{l})|\Psi _{\nu }^{(n)}>_{\mathbf{l}}=E_{\nu
}^{(n)}|\Psi _{\nu }^{(n)}>_{\mathbf{l}}  \label{33}
\end{equation}
The empty state $|\Psi _{\nu =1}^{(0)}>_{\mathbf{l}}$ with energy $E_{\nu
=1}^{(0)}=0$ defines the cluster vacuum state $|0>_{\mathbf{l}}$. There are
four singly occupied cluster states $|\Psi _{\nu }^{(1)}>_{\mathbf{l}}$ with 
$\nu =1,...,4$ which are found to be 
\begin{equation}
|\Psi _{\nu }^{(1)}>_{\mathbf{l}}=\stackrel{4}{\sum_{i=1}}\beta _{\nu i}c_{i
\mathbf{l}}^{\dagger }|0>_{\mathbf{l}}\equiv f_{\nu \mathbf{l}}^{\dagger
}|0>_{\mathbf{l}}  \label{34}
\end{equation}
where $\beta _{\nu i}$ is a ($4\times 4$)-matrix 
\begin{equation}
\beta _{\nu i}=\frac{1}{2}\left( 
\begin{array}{cccc}
1 & 1 & 1 & 1 \\ 
1 & -1 & 1 & -1 \\ 
\sqrt{2} & 0 & -\sqrt{2} & 0 \\ 
0 & \sqrt{2} & 0 & -\sqrt{2}
\end{array}
\right)   \label{35}
\end{equation}
and the corresponding eigenvalues are 
\begin{equation}
E_{1}^{\left( 1\right) }=-3t;~~~E_{2}^{\left( 1\right) }=E_{3}^{\left(
1\right) }=E_{4}^{\left( 1\right) }=t  \label{36}
\end{equation}
The eigenvectors (\ref{34}) are basis vectors of different irreducible
representations of $C_{4h}$ point group: the case $\nu =1$ describes the
lowest energy fully symmetric ($a_{g}$) solution, $\nu =2$ belongs to $b_{g}$
and $\nu =3,4$ to $e_{u}$ representation. Below we will refer to these
1-particle solutions as cluster 'orbitals'. According to (\ref{34}), for a
given cluster $\mathbf{l}$ the transformation from the original site
operators $c_{i\mathbf{l}}^{\dagger }$ ($i$=1,...,4) to the cluster
'orbital' operators $f_{\nu \mathbf{l}}^{\dagger }$ is given by the matrix 
$\beta _{\nu i}$ defined in (\ref{35}). Note that the $f_{\nu \mathbf{l}
}^{\dagger }\left( f_{\nu \mathbf{l}}\right) $ operators anticommute. The
cluster states of higher occupancy $n$=2,3,4 can be understood as a result
of successive filling of cluster 'orbitals'. For instance, for $n$=2 one
obtains six eigenstates (the cluster index $\mathbf{l}$ is dropped): 
\begin{equation}
|\Psi _{1,2,3}^{(2)}>=f_{2}^{\dagger }f_{1}^{\dagger },f_{3}^{\dagger
}f_{1}^{\dagger },f_{4}^{\dagger }f_{1}^{\dagger }|0>,~~~|\Psi
_{4,5,6}^{(2)}>=f_{3}^{\dagger }f_{2}^{\dagger },f_{4}^{\dagger
}f_{2}^{\dagger },f_{4}^{\dagger }f_{3}^{\dagger }|0>  \label{38}
\end{equation}
with corresponding energies 
\begin{equation}
E_{1,2,3}^{(2)}=-2t+V,~~~E_{4,5,6}^{(2)}=2t+V  \label{39}
\end{equation}
In the cluster 'orbital' representation, the terms $H_{t}^{(intra)}$ and 
$H_{V}^{(intra)}$ in (\ref{28}) can be now written as follows 
\begin{equation}
H_{t}^{(intra)}=\sum_{\mathbf{l}}\left( -3tf_{1\mathbf{l}}^{\dagger }f_{1
\mathbf{l}}+t\stackrel{4}{\sum_{\nu =2}}f_{\nu \mathbf{l}}^{\dagger }f_{\nu 
\mathbf{l}}\right) ,  \label{40}
\end{equation}
\begin{equation}
H_{V}^{(intra)}=\frac{V}{2}\sum_{\mathbf{l}}\stackrel{4}{\sum_{\nu =1}}
\sum_{\nu ^{\prime }\left( \neq \nu \right) }n_{\nu \mathbf{l}}n_{\nu
^{\prime }\mathbf{l}}.  \label{41}
\end{equation}
>From the result (\ref{41}), one can see that $H_{V}^{(intra)}$ is the
Hubbard term in an effective, 'multi-orbital' electronic model. Such a model
is derived below by adding to (\ref{40}) and (\ref{41}) the intercluster
hopping $H_{t}^{(inter)}$ and the Coulomb $H_{V}^{(inter)}$ terms and by
using for the latter a mean-field approximation, i.e., $n_{i,\mathbf{l}+
\mbox{\boldmath $\tau$}}n_{j\mathbf{l}}\simeq \left\langle n_{i,\mathbf{l}+
\mbox{\boldmath $\tau$}}\right\rangle n_{j\mathbf{l}}+n_{i,\mathbf{l}+
\mbox{\boldmath $\tau$}}\left\langle n_{j\mathbf{l}}\right\rangle
-\left\langle n_{i,\mathbf{l}+\mbox{\boldmath $\tau$}}\right\rangle
\left\langle n_{j\mathbf{l}}\right\rangle $. The approximated term 
$H_{V,MF}^{(inter)}$ reads 
\begin{equation}
H_{V,MF}^{(inter)}=\frac{3}{4}\left\langle N_{c}\right\rangle V\sum_{\mathbf{
l}}\stackrel{4}{\sum_{i=1}}n_{i\mathbf{l}}=\frac{3}{4}\left\langle
N_{c}\right\rangle V\sum_{\mathbf{l}}\stackrel{4}{\sum_{\nu =1}}n_{\nu 
\mathbf{l}},  \label{42}
\end{equation}
where the last equality is due to  $\sum_{i}n_{i\mathbf{l}}=\sum_{\nu
}n_{\nu \mathbf{l}}$, and $\left\langle N_{c}\right\rangle =\sum_{\nu
}\left\langle n_{\nu \mathbf{l}}\right\rangle $ is the average cluster occupancy. In the cluster
'orbital' representation, the hopping term $H_{t}^{(inter)}$ takes the
following transparent form 
\begin{equation}
H_{t}^{(inter)}=\sum_{\mathbf{l}}\sum_{\left\{ \mbox{\boldmath $\tau$}
\right\} }\sum_{\nu ,\nu ^{\prime }}T_{\nu \nu ^{\prime }}\left( 
\mbox{\boldmath $\tau$ }\right) f_{\nu \mathbf{l}+\mbox{\boldmath
$\tau$}}^{\dagger }f_{\nu ^{\prime }\mathbf{l}}.  \label{43}
\end{equation}
The ($4\times 4$)-matrices $T_{\nu \nu ^{\prime }}\left( 
\mbox{\boldmath
$\tau$ }\right) $ are related to the $M_{ij}\left( \mbox{\boldmath $\tau$}
\right) $ by a rotation 
\begin{equation}
T_{\nu \nu ^{\prime }}\left( \mbox{\boldmath $\tau$ }\right)
=-t\sum_{ij}\beta _{\nu i}M_{ij}\left( \mbox{\boldmath $\tau$
}\right) \beta _{\nu ^{\prime }j},  \label{44}
\end{equation}
because $\left( \beta ^{-1}\right) _{j\nu ^{\prime }}=\beta _{\nu ^{\prime
}j}$. Finally, by collecting the contributions (\ref{40})-(\ref{43}), we
obtain an effective 'multi-orbital' Hubbard-like Hamiltonian 
\begin{equation}
H_{t-V}=H_{t}^{(intra)}+H_{V}^{(intra)}+H_{t}^{(inter)}+H_{V,MF}^{(inter)}.
\label{45}
\end{equation}
The effective Hamiltonian (\ref{45}) has a lower symmetry compared to that
of the original Hamiltonian (1). This may lead to an artificial low-symmetry
ground-state solution, for instance, to a long-range charge/bond
ordering. Here we suggest, however, that the geometrical frustration of the
checkerboard lattice prevents this kind of long-range ordering, as was
discussed already in the beginning of this section. Therefore we supplement
the effective Hamiltonian (\ref{45}) with additional restrictions, which
prevent an artificial low-symmetry solution of (\ref{45}). We require: (a) a
homogeneous electron distribution over the lattice sites, $\left\langle
c_{i}^{\dagger }c_{i}\right\rangle =n$, with $n$ being the electron
concentration, and (b) a bond-independence of hopping amplitudes, $
\left\langle c_{i}^{\dagger }c_{j}\right\rangle =k$, where a pair of sites
$i,j$ denotes a bond ( there are six bonds within each plaquette). Note that $k
$ is a real $n$-dependent quantity to be calculated self-consistently. On
general grounds, one expects that $k>0$ in a metallic phase and $k$=0 in an
insulating phase. Both requirements (a) and (b) together impose restrictions
on cluster 'orbital' averages 
\begin{equation}
\left\langle f_{\nu \mathbf{l}}^{\dagger }f_{\nu ^{\prime }\mathbf{l}
}\right\rangle =\sum_{ij}\beta _{\nu i}\beta _{\nu ^{\prime }j}\left\langle
c_{i\mathbf{l}}^{\dagger }c_{j\mathbf{l}}\right\rangle =n\sum_{i}\beta _{\nu
i}\beta _{\nu ^{\prime }i}+k\left[ \left( \sum_{i}\beta _{\nu i}\right)
\left( \sum_{j}\beta _{\nu ^{\prime }j}\right) -\sum_{i}\beta _{\nu i}\beta
_{\nu ^{\prime }i}\right]   \label{46}
\end{equation}
By using that $\sum_{i}\beta _{\nu i}\beta _{\nu ^{\prime }i}=\delta _{\nu
\nu ^{\prime }}$ and 
\begin{equation}
\sum_{i}\beta _{\nu i}=\left\{ 
\begin{array}{c}
2;~~~\nu =1~~~~~ \\ 
0;~~~\nu =2,3,4
\end{array}
\right\}   \label{47}
\end{equation}
one obtains for $\nu =\nu ^{\prime }$ the following relations for cluster
'orbital' occupancies 
\begin{equation}
\left\langle f_{1\mathbf{l}}^{\dagger }f_{1\mathbf{l}}\right\rangle
=n+3k,~(\nu =1);~~~\left\langle f_{\nu \mathbf{l}}^{\dagger }f_{\nu \mathbf{l
}}\right\rangle =n-k,~(\nu =2,3,4).  \label{48}
\end{equation}
If $\nu \neq \nu ^{\prime }$, the expression (\ref{46}) leads to 
\begin{equation}
\left\langle f_{\nu \mathbf{l}}^{\dagger }f_{\nu ^{\prime }\mathbf{l}
}\right\rangle =0,~\left( \nu \neq \nu ^{\prime }\right) .  \label{49}
\end{equation}
For an isolated cluster with an integer electron occupancy equation (\ref{49}) is
obviously fulfilled because of symmetry arguments.

Based on the effective model (\ref{45}) we calculate the electronic band
structure from the Fourier transformation $G_{\nu \nu ^{\prime }}\left( 
\mathbf{q},\omega\right)$ of the retarded matrix Green's function:  
\begin{equation}
G_{\nu \nu ^{\prime }}\left( \mathbf{l}-\mathbf{l}^{\prime },t-t^{\prime
}\right) =\left\langle \left\langle f_{\nu \mathbf{l}}\left( t\right)
|f_{\nu ^{\prime }\mathbf{l}^{\prime }}^{\dagger }\left( t^{\prime }\right)
\right\rangle \right\rangle =-i\theta \left( t-t^{\prime }\right)
\left\langle \{ f_{\nu \mathbf{l}}\left( t\right) ,f_{\nu ^{\prime }
\mathbf{l}^{\prime }}^{\dagger }\left( t^{\prime }\right) \}
\right\rangle .  \label{50}
\end{equation}
To obtain the equation of motion for $G_{\nu \nu ^{\prime }}\left( 
\mathbf{q},\omega\right)$, we use the method\cite{yu} of
the two-time 'irreducible' Green's function. A successive differentiation of (\ref{50})
with respect to both times $t$ and $t^{\prime }$ with the use of properly
defined projection procedure lead to the Dyson's equation: 
\begin{equation}
\left[ \omega \widehat{1}-\widehat{\Omega }\left( \mathbf{q}\right) -
\widehat{\Sigma }\left( \mathbf{{q},\omega }\right) \right] \widehat{G}\left( 
\mathbf{{q},\omega }\right) =\widehat{1}.  \label{52}
\end{equation}
Here $\widehat{1}$, $\widehat{\Omega }\left( \mathbf{q}\right) $ and $
\widehat{\Sigma }\left( \mathbf{{q},\omega }\right) $ are the unit, a
frequency and a self-energy ($4\times 4$)-matrices, respectively. To define $
\widehat{\Omega }$ and $\widehat{\Sigma }$ explicitly, it is convenient to
introduce a notion of a scalar product of two fermionic operators $A$
and $B$  as 
$\left\langle A|B^{\dagger }\right\rangle =\left\langle \{ A,B^{\dagger }
\}\right\rangle $. In this notation, the matrix elements of $
\widehat{\Omega }$ are 
\begin{equation}
{\Omega }_{\nu \nu ^{\prime }}\left( \mathbf{q}\right) =\left\langle i{
\stackrel{\cdot }{f}}_{\nu \mathbf{q}}|f_{\nu ^{\prime }\mathbf{q}^{\dagger}
}\right\rangle =\left\langle \{ \left[ f_{\nu \mathbf{q}},H_{t-V}\right]
,f_{\nu ^{\prime }\mathbf{q}}^{\dagger}\}\right\rangle ,
\label{53}
\end{equation}
resulting in 
\begin{equation}
{\Omega }_{\nu \nu ^{\prime }}\left( \mathbf{q}\right) =\delta _{\nu \nu
^{\prime }}E_{\nu }^{(1)}+T_{\nu \nu ^{\prime }}\left( \mathbf{q}\right)
+\delta _{\nu \nu ^{\prime }}V\left( \sum_{\nu _{1}\left( \neq \nu \right)
}\left\langle n_{\nu _{1}}\right\rangle +\frac{3}{4}\left\langle
N_{c}\right\rangle \right) .  \label{54}
\end{equation}
Here the $E_{\nu }^{(1)}$ are given by (\ref{36}) and 
\begin{equation}
T_{\nu \nu ^{\prime }}\left( \mathbf{q}\right) =\sum_{\left\{ 
\mbox{\boldmath $\tau$}\right\} }e^{i\mathbf{{q}\cdot 
\mbox{\boldmath
$\tau$ }}}T_{\nu \nu ^{\prime }}\left( \mbox{\boldmath $\tau$
}\right) .  \label{55}
\end{equation}
The frequency matrix $\widehat{\Omega }$\ provides for a mean-field
description of the electronic band structure. In order to include effects of
electron correlations, the self-energy part $\widehat{\Sigma }$ must be
calculated.

For this purpose, the set of basis operators $f_{\nu \mathbf{q}}$ is complemented
with a new set of operators $F_{\nu \mathbf{q}}\left( \nu =1,...,4\right) $: 
\begin{equation}
F_{\nu \mathbf{q}}=i\stackrel{\cdot }{f}_{\nu \mathbf{q}}-\sum_{\nu ^{\prime
}}\Omega _{\nu \nu ^{\prime }}\left( \mathbf{q}\right) f_{\nu ^{\prime }
\mathbf{q}},  \label{56}
\end{equation}
which are orthogonal to $f_{\nu \mathbf{q}}$, i.e. $\left\langle F_{\nu 
\mathbf{q}}|f_{\nu ^{\prime }\mathbf{q}^{\prime }}^{\dagger }\right\rangle =0
$. Then, the self-energy $\widehat{\Sigma }$ has the form of an 'irreducible'
matrix Green's function\cite{yu} \ whose elements are defined by 
\begin{equation}
{\Sigma }_{\nu \nu ^{\prime }}\left( \mathbf{{q},\omega }\right) =\left\langle
\left\langle F_{\nu \mathbf{q}}|F_{\nu ^{\prime }\mathbf{q}}^{\dagger
}\right\rangle \right\rangle _{\omega }^{(irr)}=\left\langle \left\langle
F_{\nu \mathbf{q}}|F_{\nu ^{\prime }\mathbf{q}}^{\dagger }\right\rangle
\right\rangle _{\omega }-\sum_{\nu _{1}\nu _{2}}\left\langle \left\langle
F_{\nu \mathbf{q}}|f_{\nu _{1}\mathbf{q}}^{\dagger }\right\rangle
\right\rangle _{\omega }\frac{1}{\left\langle \left\langle f_{\nu _{1}
\mathbf{q}}|f_{\nu _{2}\mathbf{q}}^{\dagger }\right\rangle \right\rangle
_{\omega }}\left\langle \left\langle f_{\nu _{2}\mathbf{q}}|F_{\nu ^{\prime }
\mathbf{q}}^{\dagger }\right\rangle \right\rangle _{\omega }.  \label{57}
\end{equation}
The irreducible Green's function matrix $\left\langle \left\langle F_{\nu 
\mathbf{q}}|F_{\nu ^{\prime }\mathbf{q}}^{\dagger }\right\rangle
\right\rangle _{\omega }^{(irr)}$ obeys the following equation (the lower indices
are omitted for brevity): 
\begin{equation}
\omega \left\langle \left\langle F|F^{\dagger }\right\rangle \right\rangle
_{\omega }^{\left( irr\right) }=\left\langle F|F^{\dagger }\right\rangle
+\left\{ \left\langle i\stackrel{\cdot }{F}|F^{\dagger }\right\rangle
+\left\langle \left\langle i\stackrel{\cdot }{F}|-i\stackrel{\cdot }{F}
^{\dagger }\right\rangle \right\rangle _{\omega }^{\left( irr\right)
}\right\} \frac{1}{\left\langle F|F^{\dagger }\right\rangle }\left\langle
\left\langle F|F^{\dagger }\right\rangle \right\rangle _{\omega }^{\left(
irr\right) }  \label{571}
\end{equation}
We truncate this equation by neglecting in (\ref{571}) the next-order
'irreducible' matrix Green's function $\left\langle \left\langle i\stackrel{
\cdot }{F}|-i\stackrel{\cdot }{F}^{\dagger }\right\rangle \right\rangle
_{\omega }^{(irr)}$. For the effective 'multiorbital' model (\ref{45}), the
truncation corresponds to the first step of the Hubbard-I approximation and
results in the following form of $\widehat{\Sigma }$: 
\begin{equation}
{\Sigma }_{\nu \nu ^{\prime }}\left( \mathbf{{q},\omega }\right) =\left[
\omega \widehat{1}-\widehat{\Lambda }\left( \mathbf{q}\right) \right] _{\nu
\nu ^{\prime }}^{-1}\left\langle F_{\nu ^{\prime }\mathbf{q}}|F_{\nu
^{\prime }\mathbf{q}}^{\dagger }\right\rangle ,  \label{58}
\end{equation}
where the frequency matrix $\widehat{\Lambda }\left( \mathbf{q}\right)
$ is given by  
\begin{equation}
\Lambda _{\nu \nu ^{\prime }}\left( \mathbf{q}\right) =\frac{\left\langle i
\stackrel{\cdot }{F}_{\nu \mathbf{q}}|F_{\nu ^{\prime }\mathbf{q}}^{\dagger
}\right\rangle }{\left\langle F_{\nu ^{\prime }\mathbf{q}}|F_{\nu ^{\prime }
\mathbf{q}}^{\dagger }\right\rangle }.  \label{59}
\end{equation}
In the Hubbard-I approximation applied to the standard Hubbard model, the
corresponding self-energy is $\mathbf{q}$-independent. We shall apply a
similar approximation here by dropping the intercluster hopping term $
H_{t}^{\left( inter\right) }$ which appears in the equation $i\stackrel{\cdot }{F}_{\nu 
\mathbf{q}}=\left[ F_{\nu \mathbf{q}},H_{t-V}\right] $. In the following, $
\widehat{\Lambda }$ and $\widehat{\Sigma }$ are $\mathbf{q}$-independent
matrices and the latter is of the form 
\begin{equation}
{\Sigma }_{\nu \nu ^{\prime }}\left( \omega \right) =\delta _{\nu \nu ^{\prime
}}\frac{\left\langle F_{\nu \mathbf{l}}|F_{\nu \mathbf{l}}^{\dagger
}\right\rangle }{\omega -\left\langle i\stackrel{\cdot }{F}_{\nu \mathbf{l}
}|F_{\nu \mathbf{l}}^{\dagger }\right\rangle /\left\langle F_{\nu \mathbf{l}
}|F_{\nu \mathbf{l}}^{\dagger }\right\rangle }.  \label{60}
\end{equation}
With the definition (\ref{56}), one obtains explicitly 
\begin{equation}
F_{\nu \mathbf{l}}=V\delta N_{\mathbf{l}}^{\nu }f_{\nu \mathbf{l}},
\label{62}
\end{equation}
where 
\begin{equation}
\delta N_{\mathbf{l}}^{\nu }=N_{\mathbf{l}}^{\nu }-\left\langle N_{\mathbf{l}
}^{\nu }\right\rangle ,~~~N_{\mathbf{l}}^{\nu }=N_{\mathbf{l}}-n_{\nu 
\mathbf{l}}=\sum_{\nu _{1}\left( \neq \nu \right) }n_{\nu _{1}\mathbf{l}}.
\label{63}
\end{equation}

\begin{figure}[tbp]
\includegraphics[clip=true,width=100mm]{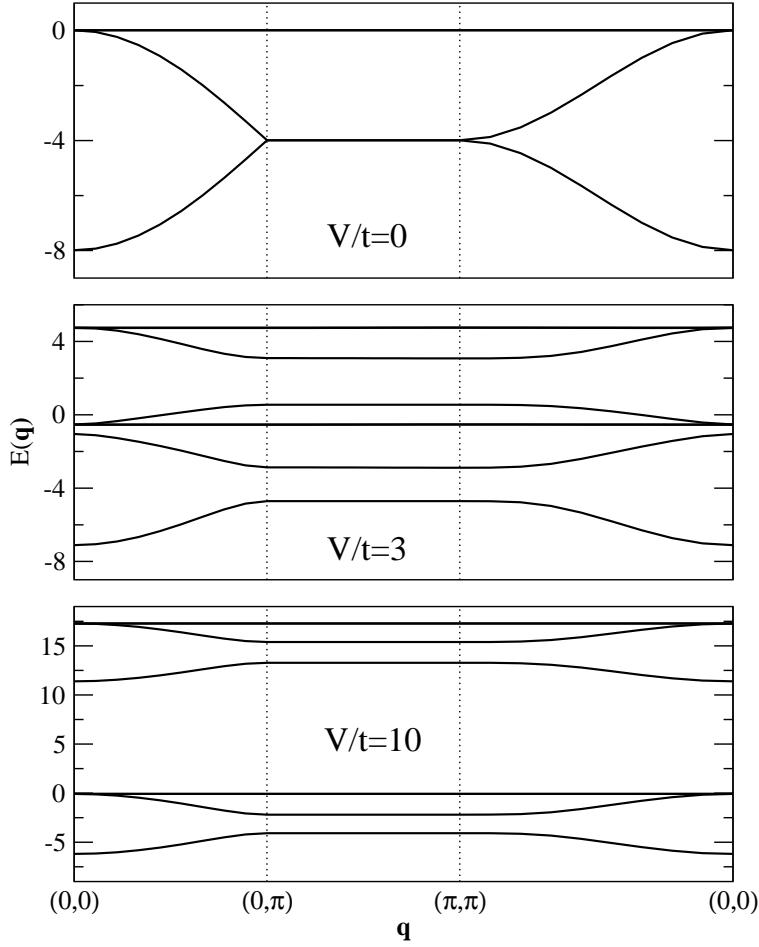}
\caption{Energy bands of spinless fermions on the checkerboard lattice
calculated in the cluster approach for different values of inter-site
repulsion $V/t=0,3<V_c/t$ and $V/t=10>V_c/t \simeq 4$. The bands are 
unfolded along the highly symmetrical directions in the square Brillouin zone; 
energy unit is t $\equiv $ 1. The electron concentration is chosen as 
$n = \frac{1}{2}$ (equivalent to $\la N_{c}\ra$ =2) and the chemical potential 
$\protect\mu $ is fixed at zero energy.}
\end{figure}
Then the diagonal elements of the frequency matrix entering into the
denominator of (\ref{60}) are found to be 
\begin{equation}
\Lambda _{\nu \nu }=\frac{\left\langle i\stackrel{\cdot }{F}_{\nu \mathbf{l}
}|F_{\nu \mathbf{l}}^{\dagger }\right\rangle }{\left\langle F_{\nu \mathbf{l}
}|F_{\nu \mathbf{l}}^{\dagger }\right\rangle }=E_{\nu }^{\left( 1\right) }+
\frac{3}{4}V\left\langle N_{c}\right\rangle +V\frac{\left\langle N_{\mathbf{l
}}^{\nu }\left( \delta N_{\mathbf{l}}^{\nu }\right) ^{2}\right\rangle }{
\left\langle \left( \delta N_{\mathbf{l}}^{\nu }\right) ^{2}\right\rangle }.
\label{64}
\end{equation}
Both averages, $\left\langle \left( \delta N_{\mathbf{l}}^{\nu }\right)
^{2}\right\rangle $ and $\left\langle N_{\mathbf{l}}^{\nu }\left( \delta N_{
\mathbf{l}}^{\nu }\right) ^{2}\right\rangle $, are approximated in a
mean-field manner. For instance, in this approximation, the mean value $
\left\langle \left( \delta N_{\mathbf{l}}^{\nu }\right) ^{2}\right\rangle $
which describes the intracluster charge correlations, reads 
\[
\left\langle \left( \delta N_{\mathbf{l}}\right) ^{2}\right\rangle
=\sum_{\nu _{1}\left( \neq \nu \right) }\left\langle n_{\nu _{1}\mathbf{l}
}\right\rangle \left( 1-\left\langle n_{\nu _{1}\mathbf{l}}\right\rangle
\right) .
\]
We summarize the results for the self-energy as follow
 $\left( \left\langle n_{\nu 
\mathbf{l}}\right\rangle =\left\langle n_{\nu }\right\rangle \right) $:
 
\begin{equation}
{\Sigma }_{\nu \nu }\left( \omega \right) =\frac{\sum_{\nu _{1}\left( \neq \nu
\right) }\left\langle n_{\nu _{1}}\right\rangle \left( 1-\left\langle n_{\nu
_{1}}\right\rangle \right) }{\omega -\Lambda _{\nu \nu }},  \label{65}
\end{equation}
\begin{equation}
\widehat{\Lambda }\left( \omega \right) =E_{\nu }^{(1)}+\frac{3}{4}
V\left\langle N_{c}\right\rangle +V\frac{\sum_{\nu _{1}\left( \neq \nu
\right) }\left\langle n_{\nu _{1}}\right\rangle \left( 1-\left\langle n_{\nu
_{1}}\right\rangle \right) ^{2}}{\sum_{\nu _{1}\left( \neq \nu \right)
}\left\langle n_{\nu _{1}}\right\rangle \left( 1-\left\langle n_{\nu
_{1}}\right\rangle \right) }+V\frac{\sum_{\nu _{1}\left( \neq \nu \right)
}\sum_{\nu _{2}\left( \neq \nu ,\nu _{1}\right) }\left\langle n_{\nu
_{1}}\right\rangle \left( 1-\left\langle n_{\nu _{1}}\right\rangle \right)
\left\langle n_{\nu _{2}}\right\rangle }{\sum_{\nu _{1}\left( \neq \nu
\right) }\left\langle n_{\nu _{1}}\right\rangle \left( 1-\left\langle n_{\nu
_{1}}\right\rangle \right) }.  \label{66}
\end{equation}
In this expression the second term is the inter-cluster Hartree-Fock
correction. The third and fourth terms are intra-cluster Hubbard-I type
correlation corrections, where the latter appears only for the present
'multi-orbital' Hubbard Hamiltionian. In the common Hubbard model (with just
one 'orbital') only the third correction would be present. To perform
self-consistent band-structure calculations, a chemical potential $\mu $ is
introduced in a standard way. For a given spinless-fermion concentration $n$
varying within the range $0\leq n\leq 1$, the value of $\mu $ is determined
from the equation 
\begin{equation}
n=\frac{1}{4}\left\langle N_{c}\right\rangle =\frac{1}{4}\stackrel{4}{
\sum_{\nu =1}}\left\langle n_{\nu }\right\rangle ,  \label{67}
\end{equation}
where $\left\langle n_{\nu }\right\rangle $ is an average 'orbital' cluster
occupancy 
\begin{equation}
\left\langle n_{\nu }\right\rangle =\frac{1}{M}\sum_{\mathbf{q}
}\int_{-\infty }^{\mu }d\omega \left( -1/\pi \right) \mathop{\rm Im}G_{\nu
\nu }\left( \mathbf{{q},\omega }\right) .  \label{68}
\end{equation}
Here the summation is over $M=L/4$ $\mathbf{q}$-vectors in the first
Brillouin zone of the square lattice formed by the basic clusters in the
checkerboard lattice. The 'orbital' occupancies should obey the relations 
(\ref{48}).

The most representative results of the self-consistent band structure
calculations are shown in Figs. 6 and 7. In Fig. 6, band spectra for
different values of coupling $V$ are given along highly symmetrical
directions in the square Brillouin zone. To show the insulating gap opening
in the band spectrum with increasing $V$, the electron concentration $n$=1/2
corresponding to an integer mean cluster occupancy $\left\langle N_c
\right\rangle $=2 is chosen; the chemical potential $\mu$ is located at zero
energy. From the upper panel, a doubly degenerate completely flat branch on
the top of two highly dispersive bands is seen for the non-interacting case, 
$V$=0. For finite coupling $V$, but less than some critical value $V< V_c$,
these four branches are split and the chemical potential intersects one of
the dispersive branches as shown in the middle panel of Fig. 6. In this
weakly correlated regime of the model one expects a metallic state of the
system. We found that starting from $V= V_c \approx 4t$, the spectrum is
clearly split into the low- and high-energy Hubbard subbands separated with
a gap growing as $V$ increases above $V_c$. For a given concentration 
$\left\langle N_c \right\rangle $=2 and $V> V_c$, the four lowest Hubbard
subbands are filled completely and located below the chemical potential $\mu$
as shown for $V=10$ in the lower panel of Fig. 6. If one includes a weak
next neighbor hopping $t^{\prime}$, the former flat energy branches acquire
a small dispersion, the picture of the insulating gap opening still retains
at slightly changed value of the critical coupling $V_c$.

\begin{figure}[tbp]
\includegraphics[clip=true,width=90mm,angle=90]{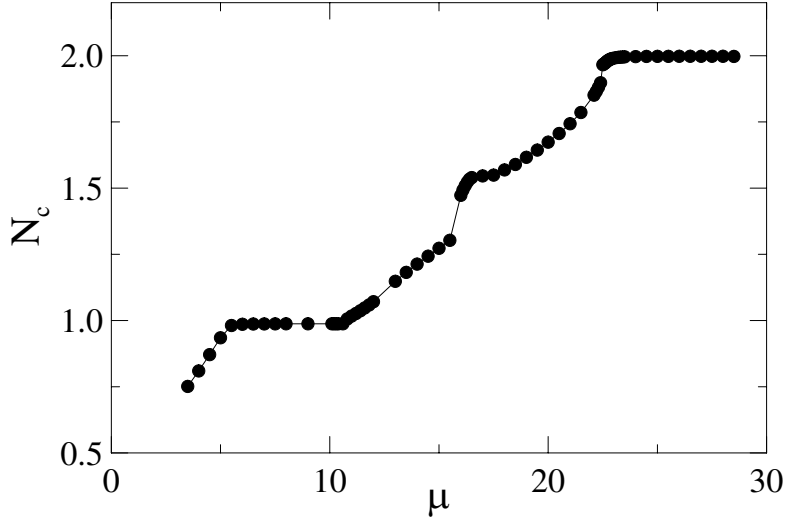}
\caption{Mean cluster occupancy $\la N_c\ra$ calculated as a function of the
chemical potential $\protect\mu$ at V/t = 10. Two wide flat regions with $d 
\la N_c\ra/d\protect\mu$ = 0 at $\la N_c\ra$ =1,2 indicate an insulating
state of the system.}
\end{figure}

In this paper, however, we restrict ourselves to the study of the limit, 
$t^{\prime}$=0. In this limit, to avoid discussing a rather special case of
the partially filled upper flat band, the electron concentration is chosen
to vary in the range $n \leq 1/2$. Within this range, for strong coupling $V$
the model displays an insulating state at the integer cluster occupancy 
$\left\langle N_c \right\rangle $=1 as well. This can be easily seen from
Fig. 7, where the calculated mean cluster occupancy $\left\langle N_c
\right\rangle$ is depicted as a function of the chemical potential $\mu$ for 
$V$=10. The charge compressibility $d\left\langle N_c \right\rangle/d\mu$ is
found to be zero in a wide range of varying $\mu$ at both integer
occupancies $\left\langle N_c \right\rangle $=1,2. Near the occupancy value 
$\left\langle N_c \right\rangle $=1.5, the sharp change of $\left\langle N_c
\right\rangle $ is connected to the fact that $\mu$ intersects the flat band
peculiarity in the density of states.

\section{Conclusion}

In this work we have studied an extended Hubbard type spinless fermion model
for the frustrated checkerboard lattice which is a 2D analogue for
pyrochlore or spinel-type lattices. We have studied the possibility of a
metal-insulator phase transition as function of the inter-site Coulomb
interaction and the band filling. We have used single-site and cluster
representations of the model. In the latter the intra-cluster correlations
are accounted for exactly by transforming to a 'cluster orbital' basis.
Within a Green's function approach decouplings on Hartree-Fock and
'Hubbard-I' type level for the inter-site Coulomb term have been employed.
In both approaches we find a metal-insulator transition for increasing V for
half filling (that is one fermion per two sites) and in the cluster
approach even for quarter filling. The MI transition is of the Mott
Hubbard-type and is associated with a gap opening in the quasiparticle
excitations. The critical interaction for the MI transition is V$_{c1}$/t=4.86 
in the single site approach and $V_{c}/t\simeq$4 in the cluster
approach are in reasonable agreement. We also consider the possibility of
charge ordering within the single-site approach where we find a transition
to a staggered CO state with $\bf Q$ = (0,0) at V$_{c2}$/t=6.47. This state 
may be a result of the approximations employed since for
V/t $\rightarrow\infty$ CO has to vanish due to the macroscopic degeneracy of
the ground state. Stabilisation of CO may indeed require the coupling to the
lattice to lift this degeneracy \cite{Zhang1}. Further progress in
understanding the nature of the isulating state caused by the \emph{inter-site} 
Coulomb interaction may require the use of more advanced methods
like cluster dynamical mean field methods which has sofar not been achieved. 

The authors are grateful to Prof. P. Fulde for suggesting the subject of studies and valuable discussions and comments.


\begin{thebibliography}{99}
\bibitem{note}  Note that the ground state at $V/t \rightarrow \infty$ is quite different from that at $V/t=0$ although CO is absent in
both cases. For example the expectation values of the nearest- and next-nearest-neighbor charge correlation are different.
\end{thebibliography}
\end{document}